\documentclass[]{article}
\usepackage{graphicx}

\usepackage{booktabs}

\usepackage{amssymb}
\usepackage{amsmath}
\usepackage{graphicx}

\usepackage[title]{appendix}

\usepackage[citestyle=authoryear,backend=biber]{biblatex}
\usepackage{subfigure}
\usepackage{gensymb}
\usepackage{textcomp}

\usepackage{url}
\usepackage[utf8]{inputenc}
\usepackage[T1]{fontenc}
\usepackage{hyperref}

\usepackage[english]{babel}
\DeclareUnicodeCharacter{0301}{\'{e}}

\usepackage[marginparwidth=2.5cm]{geometry}
\usepackage{xargs} % Use more than one optional parameter in a new commands
\usepackage[pdftex,dvipsnames]{xcolor}  % Coloured text etc.
\usepackage[colorinlistoftodos,prependcaption,textsize=tiny]{todonotes}
\newcommandx{\dk}[2][1=]{\todo[linecolor=red,backgroundcolor=red!25,bordercolor=red,#1]{#2}}
\newcommandx{\vs}[2][1=]{\todo[linecolor=blue,backgroundcolor=blue!25,bordercolor=blue,#1]{#2}}
\newcommandx{\am}[2][1=]{\todo[linecolor=OliveGreen,backgroundcolor=OliveGreen!25,bordercolor=OliveGreen,#1]{#2}}

\usepackage{authblk}
\usepackage{blindtext}

\usepackage{nameref}

\usepackage{tikz}
\usepackage{tkz-berge}
\usetikzlibrary{positioning}

\providecommand{\keywords}[1]{\textbf{\textit{Keywords:}} #1}

\title{Latent Dirichlet Allocation Models for World Trade Analysis}

\author[1]{Diego Kozlowski\thanks{diego.kozlowski@uni.lu}}
\author[2]{Viktoriya Semeshenko}
\author[2]{Andrea Molinari}
\affil[1]{DRIVEN, FSTM, University of Luxembourg, Luxembourg}
\affil[2]{Universidad de Buenos Aires. Facultad de Ciencias Econ\'omicas. Buenos Aires, Argentina. CONICET-Universidad de Buenos Aires. Instituto Interdisciplinario de Econom\'ia Pol\'itica  de Buenos Aires. Buenos Aires, Argentina}
\date{}                     %% if you don't need date to appear

\begin{document}

\maketitle

\begin{abstract}

The international trade is one of the classic areas of study in economics. 
Nowadays, given the availability of data, the tools used for the analysis can be complemented and enriched with new methodologies and techniques that go beyond the traditional approach. The present paper shows the application of the Latent Dirichlet Allocation Models, a well known technique from the area of Natural Language Processing, to search for latent dimensions in the product space of international trade, and their distribution across countries over time. We apply this technique to a dataset of countries' exports of goods from 1962 to 2016.  The findings show the possibility to generate higher level classifications of goods based on the empirical evidence, and also allow to study the distribution of those classifications within countries. The latter show interesting insights about countries' trade specialisation.

\end{abstract}

\section{Introduction}
\label{sec:intro}

The role that countries play in the global market is profoundly determined by their insertion into global value chains, and by the types of goods they produce for the global market \parencite{coe2004globalizing, gereffi2005governance,gereffi1994organization}. 

Production systems, which were traditionally analyzed as almost independent national systems, are now continuously more connected on a global scale. 
Due to the increasingly complex and interconnected nature of global supply chain networks, a recent strand of research has applied network science methods to model global supply chain growth and subsequently analyse various topological features of these structures.
Obviously, this depends on the dataset in use, as it defines the topology of the network.

In recent years, we have been witnessing a continuous  growth of available data. This situation also poses a great challenge, namely, how to extract hidden relations, determine appropriate patterns, clusters and trends to extract valuable conclusions from such large volumes of data \parencite{Padhy2012arXiv}.   

Traditional analysis tools are incapable to handle such complexity alone because it requires time and efforts to extract and analyse information. 
On the other hand, interdisciplinary sciences provide different techniques and tools to apply to the analysis of this volume of data. 
The application of network formalism in the field of socioeconomic science has experienced unprecedented growth in recent decades \parencite{Barabasi2011Nature, Caldarelli2007Scale, Ermann2012, Fagiolo2011}. 
Also, there is a wide literature that studies international trade at the product level \parencite{balassa1965trade,lall2000technological,lall2006sophistication, haveman2004alternative}. In particular, these connections can be analyzed as a bipartite graph among countries and products \parencite{Guan2018, straka2017grand, ferreira2016topology, caldarelli2012network}, and the complexity of production can be explored using the product space \parencite{Hidalgo2009a, Hidalgo2009, Hidalgo2007}.
The world trade network can also be examined using multiplex and multilayer networks \parencite{Battiston2014PhysRev, Kivela2011Complex, Alves2018}. 

In this paper, we adopt a different approach to extract interesting and significant patterns from bilateral trade data, using the Latent Dirichlet Allocation (LDA) modelling technique \parencite{blei2003latent}.  
Topic models have emerged as an effective method for discovering useful structure in data. At the same time, LDA is a statistical approach used in topic modeling for discovering hidden topics in large corpora of text. 

Recently, a growing number of researchers are beginning to integrate topic models into various datasets \parencite{pritchard2000inference, Rosa2015ProbabilisticTM, 1467486, 5346483, Hu2009APT}, not only for document collections.
To the best of our knowledge, our work is the first effort to adapt and apply this technique for countries' exports.

We find very suitable an analogy between topic modeling in texts and trade. In our adaptation of LDA, a set of countries plays the role of text documents, products play the role of words, and components (i.e. latent dimensions within which these products group) play the role of topics. Based on the model of \cite{blei2003latent}, we suggest a generative process to detect these latent dimensions in the product space and build an alternative trade nomenclature directly from data. 
Then, using these latent dimensions, we analyze those components' participation within countries' export baskets.

Our main contributions and results can be summarized as follows: we develop a generative model, based on a well established methodology usually  used in the field of Natural Language Processing, to study the international trade flows.
This model looks for automatic grouping of the products in latent components. We study these latent components, characterizing each by type of production, complexity and its relation to a specific country over time. Then, we use the components to briefly characterize the role in global trade of different groups of countries. The results that emerge from our model are in line with the specialized economic and trade history literature.

The paper is organized in the following way: in section  \ref{sec:data} we describe the dataset in use, in section \ref{sec:method} we introduce the notations and explain the methodology applied in the model, in section \ref{sec:results} we present the obtained results, and in section \ref{sec:discussion} we conclude. 

\section{Data} 
\label{sec:data}

To apply the LDA technique, we used the United Nation Commodity Trade Statistics Database (COMTRADE) dataset\footnote{This dataset has been extracted on March, 2019.} of each country's (four-digits) disaggregated exports from the Center for International Development at Harvard University. Such dataset contains trade data for around 250 countries and territories, and takes the raw trade data on goods from countries’ reporting to the United Nations Statistical Division (COMTRADE). 

We used these data instead of the raw COMTRADE statistics because such data may contain some inconsistencies. To address this issue, the Center for International Development uses the Bustos-Yildirim Method to clean data and "account for inconsistent reporting practices and thereby generate estimates of trade flows between countries".\footnote{See \url{https://atlas.cid.harvard.edu/about-data} for more details.} Such method assumes that since these data are recorded both as exports and as imports, cross-referencing countries' reported trade flows against each other can produce reliable estimations. It consists of first correcting bilateral import values and then comparing them to the reverse flows reported by the exporting partner.\footnote{Imports are reported CIF (i.e. including freight and insurance costs) and exports free on board (FOB).} Their (per-country) estimated index of reliability for reporting trade flows measures the consistency of trade totals reported by all exporter and importer combinations over time. Finally, they generate their own trade values' estimates using the data reported by countries together with such reliability index.

Bilateral trade flows are mainly recorded in two trade classification systems: Harmonized System (HS) and Standard International Trade Classification (SITC), and data presents four dimensions: exporter, importer, product, and year.
While both classifications are valid, there is a "time versus disaggregation" trade-off entangled in the decision of which dataset to use. SITC data has a longer time-series (1962-2016) but it covers fewer goods (i.e. at higher levels of aggregation, up to 4-digits, approximately 750 products). On the other hand, HS data, being a newer classification, offers a more contemporary and detailed classification of goods (i.e. disaggregated up to 6-digits, with approximately 5,000 goods), but with the downside of offering a relatively shorter time period (1995-2017).

We chose to work with SITC (in this case Revision 2) in order to have a larger time series, having slightly more aggregated data (i.e. 4- instead of 6-digits) \parencite{DataRev2}. Moreover, we reckon that 750 products should be enough to allow us to apply the LDA technique, as it should allow for enough (but not too much) granularity when labelling the components. For such dataset, we make an empirical search for the best number of latent dimensions.

\section{Methodology}
\label{sec:method}

In this section we describe a probabilistic model constructed to study the trade flow data with the aim to generate an automatic grouping of the products. 

This cannot be achieved using traditional clustering techniques in high dimensional spaces \parencite{aggarwal2001surprising}, due to the fact that a product can be used or consumed as an intermediate and/or final product at the same time, which means that groups can not be  exclusive  \parencite{molinari2016especializacion}. Therefore, the problem we are dealing with can be examined with \textit{fuzzy} clustering. 

At the same time, we need to deal with mitigating high-dimensional data issues through dimensionality reduction. This is possible  due to the fact that we can explode similarities between the products. The dimension of the problem of grouping the products can be thought of as a $\mathcal{R}^{N*P*Y}$ space. That is, the interaction of $N$ countries, $P$ products and $Y$ years.

We find it appropriate to use LDA to group products. While \cite{blei2003latent} look for a latent dimension of $k$ \textit{topics}, embedded in a highly dimensional dictionary distributed over the texts that compose the corpus, here we are looking for a latent dimension of $k$ \textit{components}, embedded in a highly dimensional classification of products distributed along the countries over the years. 

We use the following terms to define our probabilistic topic model: 

\begin{itemize}
\item
\textbf{\emph{product}} is a \emph{basic discrete unit of analysis}, defined as an item in a classification (SITC). We represent products using unit-basis vectors, where the superscript \(i\) stands for the $i^{th}$ product in the classification and the $i^{th}$ element in the vector. The $V^{th}$ product of the classification is the vector  \(w\), such that \(w^v\)=1 and \(w^u\)=0, \(u\neq v\).
\item
\textbf{\emph{country-year}} is a sequence of \textbf{N} products, defined as \(W= (w_1, w_2, ..., w_N)\).
\item
\textbf{corpus} is the collection of \textbf{M} country-years, defined as \(D = (d_1, d_2,..., d_M)\).
\item
\textbf{\emph{component}} is a latent dimension on the corpus, defined as \emph{K}.
\end{itemize}

The objective behind the classification of the products is twofold: on the one hand, look for a distribution of components over each country-year; on the other, analyse the distribution of the products within each of the components.

\subsection{Generative process}

In the original model proposed by \cite{blei2003latent}, the words are supposed to be random realisations of chained distributions, ignoring the order in which the words appear in the document. Even when we know that the real data generating process is far from what our model proposes, this inference process can still provide useful insight on the latent dimensions we are looking for. The basic idea of the generative process is that, given the amount of dollars exported by a country in a specific year, the assignment of the product that will be exported comes from a random mixtures over latent components, where each component is characterised by a distribution over products. 
The sequence of the data generation can be described as follows:

\begin{itemize}
\item For each country-year in the corpus, we assume that exports come from a following two-stage process:

\begin{itemize}
	\item
	choose randomly a distribution for the components,
	\item for every dollar exported:
	
	\begin{itemize}
	    \item choose randomly the component to which it belongs, and
		\item choose randomly a product from the distribution corresponding to that component.
	\end{itemize}
\end{itemize}

\end{itemize}

The data generating process can be formalised as follows:

\begin{enumerate}
\def\labelenumi{\arabic{enumi}.}
\item
For every component $k \in \{1,2,... K\}$

\begin{itemize}
\item Generate a distribution over the products
$\beta_k \sim Dir(\eta)$, where $\eta \in \mathcal{R}_{>0}$   is fixed\footnote{In this case $\eta= 1/K$.}
\end{itemize}

\item
For each country-year $d \in \{1,2,... D\}$

\begin{itemize}
\item
Generate a vector of component proportions
$\theta_d \sim Dir(\alpha)$, where $\alpha \in \mathcal{R}_{>0}$
is fixed\footnote{In this case $\alpha=1/K$.}
\item

For every exported dollar:

\begin{enumerate}
	\def\labelenumi{\roman{enumi}.}
	\item
	generate an allocation of the component $z_{dn} \sim Mult(\theta_d)$
	\item
	assign the product $w_{dn} \sim Mult(\beta_{zn})$
\end{enumerate}
\end{itemize}
\end{enumerate} 

A Dirichlet process is a family of stochastic processes where the realizations are themselves probability distributions. 
It is often used in Bayesian inference to describe the prior knowledge about the distribution of random variables—how likely it is that the random variables are distributed according to one or another particular distribution.

The parameters defining the Dirichlet distribution (here,  $\eta$ and $\alpha$) determine the degree of concentration of the resulting distributions. 
For a $Dir(\alpha)$ distribution, $\alpha$ defines the degree of symmetry of the multinomial distributions that the process generates. 
With values much smaller than 1, the resulting distributions will be highly concentrated on some elements, while values much larger than 1 would generate very uniform distributions. 
In terms of our problem, $\alpha$ controls the mixture of components for any given country, and parameter $\eta$ controls the distribution of products per component. A very small $\alpha$ will generate that each country has few characteristic components, while a very small $\eta$ will generate a very asymmetric distribution over the products, and therefore there will be a few very important products, and the rest with almost null probability.

\section{Results}
\label{sec:results}

In this section we present results for the analysis of components, first at their distribution and then at their main country over time. We confine our analysis over the period 1962-2016, for the 250 reporting countries and P products (goods, not services), which, as mentioned, in SITC Rev.2 \parencite{DataRev2} at 4-digits are approximately 750. In other words, we work within an order of magnitude similar to that of a regular dataset in a traditional Topic Modelling problem.  As mentioned before, prioritising a longer time series, we decided to use the SITC (Rev. 2) 4-digit nomenclature.
 
In the following sections, results are discussed in two stages. We first walk the reader through the decision of the number of components, also discussing the labelling process adopted, and then analyse the evolution of exports in the main country for each component.

\subsection{Analysis of Components}

In this subsection, we first explain the (granularity vs. economic interpretation) trade-off faced when using trade data with LDA. We then describe the process of finding the best suitable number of components ($k$) for our problem, and the labelling of each component, to conclude with some reflections about the findings for the chosen $k$.

Hyperparameter $k$ stands for the amount of components and plays a fundamental role in the model. Fewer components (i.e. small $k$) will tend to reflect broader concepts. On the other hand, if $k$ is larger than the cardinality of the latent space (i.e. the implicit space for the grouped products is smaller than the number of proposed components),\footnote{In the limit, in our case, one could have one topic per product.} this can generate repeated or over-specific components. In other words, in our case this issue poses a trade-off between granularity and well-defined (i.e. easily "taggable") components.

First, we ran the model for various values of $k=2, 4, 6, 8, 10, 20, 30, 40, 50, 100, 200$. A first result, observed for all these values of $k$, is that components that group best (for every $k$) are those containing: petroleum and derivatives, electronics, machinery and textiles. As mentioned above, hyperparameter $k$ defines the components' specificity. However, those phenomena worth exploring (and for economic interpretation) can be found at different levels of granularity. Hence, a first problem observed when analysing the different exercises is to define a suitable granularity for the components.  For relatively low values of $k$ (i.e. up to $k=10$) the petroleum component always stands out. Conversely, from $k=20$ we also find other sectors (e.g. electronic products, textiles, etc) in some components, while others hold a mixture of products that is harder to rationalise as a latent dimension. For values of $k$ between 20 and 50 the resulting composition for each component is rather stable, resulting in a good balance between more easily interpretable (i.e. taggable) components, together with an interesting level of granularity. For values of $k$, higher than 50, components tend to repeat themselves.

\begin{figure}[h]
	\centering	
	\includegraphics[width=0.65\linewidth]{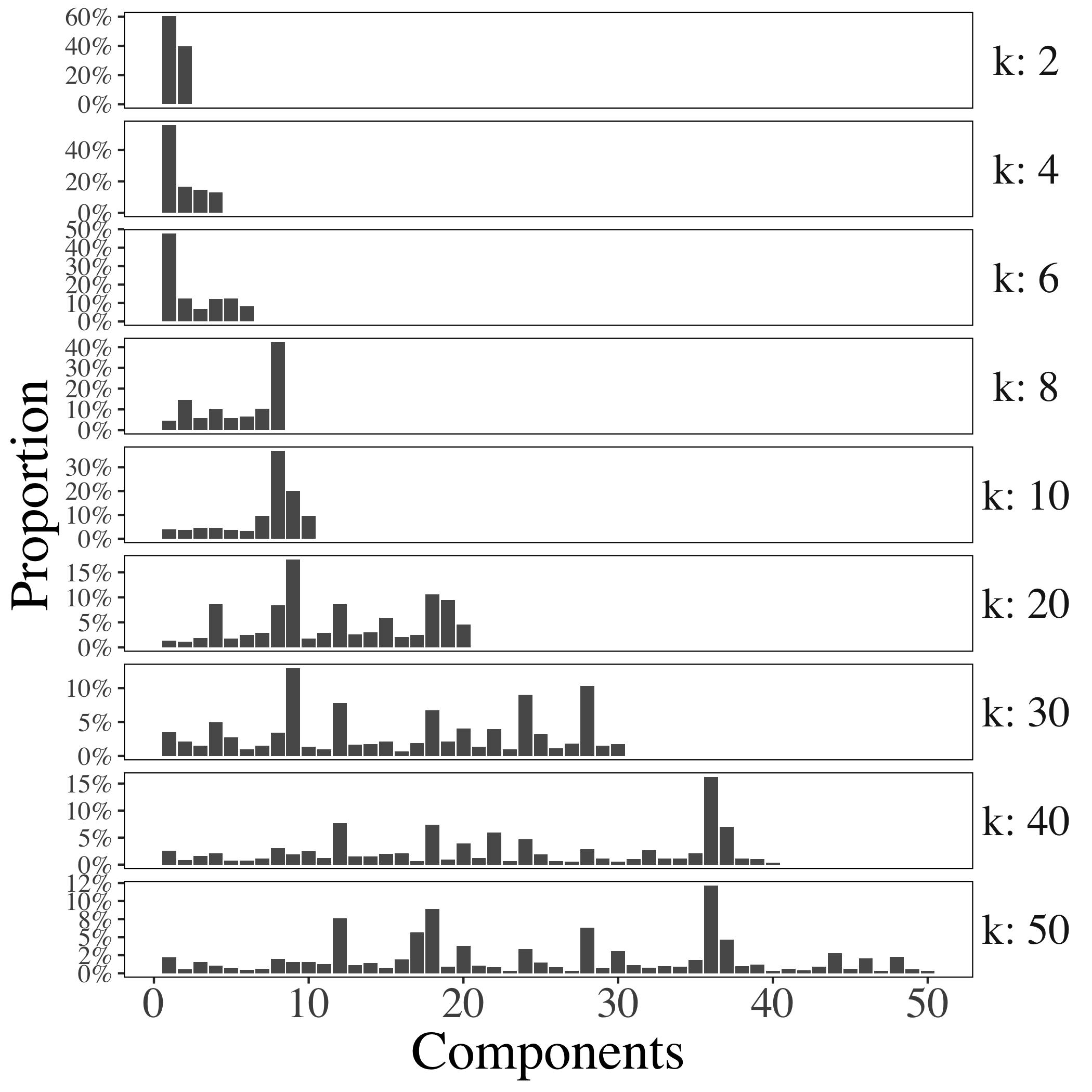}
	\caption{Average proportion of the components, various values of k.}
	\label{fig:componentes}
\end{figure}

Figure \ref{fig:componentes} shows an irregular distribution of components (for $k=2$ to $k=50$).\footnote{This does not weight components by total exports, i.e. shows an equal basis among countries.} Some components stand out, such as the first component for $k=2,4,6$; or the eighth component in $k=8,10,20$. These components are mainly composed of petroleum products. This result can be interpreted in two (not mutually exclusive) ways. It could reflect that the dataset contains relatively more primary producing countries than industrial manufacturing ones, but it can also be showing that the former countries' exports are more concentrated on primary products.

A second step consisted of choosing a $k$ that was appropriate, considering such granularity versus taggability trade-off. Since there is no clear optimum value for $k$, and although the literature within the text analysis domain has contributed with some proposals, this parameter should come from a substantive search where the topics (or components) found are closer to the object of study that is analyzed \parencite{bonilla2013elevated, quinn2010analyze}.

To explore the distribution of products over components, and their cumulative function, we developed a dynamic dashboard.\footnote{See \url{https://diego-kozlowski.shinyapps.io/LDA_worldtrade/}. Appendix \ref{appendix:k2} shows the step-by-step labelling exercise and some possible economic interpretation of results for  $k=2$} Such distribution is plotted according to a widely used technological exports classification \parencite{lall2000technological}. After a substantive search (which involved the mentioned estimation, comparison and manual exploration of the model with different $k$), we feel comfortable with choosing $k=30$ as the number of components offering the best trade-off between having enough (economically interpretable) granularity and (a relatively low) components' repetition.

For $k=30$, we also tested for different $\eta$. As we want our components to have an asymmetric distribution, to facilitate their labelling, we ran the model with small values of $\eta$. Specifically, we tested the model for $\eta=1/30,1/60,1/90,1/120$. Components' composition did not show substantive changes with different values for $\eta$, suggesting that the model is robust to variations in the priors. For this reason, and given that the default value $\alpha=\frac{1}{k}$ gives good results in terms of countries specialisation, we decided to keep the default values of $\alpha=\eta=\frac{1}{k}$ for the mentioned various runs with different values of $k$.

In a third and final step, we \emph{manually} labelled the components for the chosen model ($k=30$). The mentioned dynamic dashboard (with each component product composition and the distribution is plotted according to a widely used technological exports classification) helped the labelling exercise in terms of re-grouping products within each component by their technological complexity. Frequently, and even in text topic modeling, component labeling is quite difficult due to the lack of a generalisation criteria. We found that a downside of the LDA technique within the trade flows domain is that this issue is reinforced, since the subjective search for a comprehensive concept of products traded among countries can turn to be a more complex task than searching for a general concept over a group of words. On the upside, polysemy, an frequent problem found in texts, does not exist when using trade data, where all signifiers (classification indexes) refer to a single and unambiguous meaning. However, other new problems arise, e.g. deciding upon the trade nomenclature or the data disaggregation level (which could be associated with choosing the language of the corpus in text analysis). In our model, we first observed that the usual practice of looking at the first ten elements of the distribution was not sufficient to find a general label for each component, and for this reason we develop a more comprehensive dashboard.

Table \ref{table:desc_comp} shows the labels for our model (with $k=30$), with a general description by component, except when that is not possible (e.g. component 19), together with a `subgroup' that allows for a finer (or more detailed) product specification and, in the case of industrial products, the level of technological complexity (according to \cite{lall2000technological}). Finally, the last column displays the country for which each component has the highest share (taking an average over the whole time period).

% \usepackage{booktabs}
\begin{table}[htbp]
\begin{tabular}{@{}lllll@{}}
\toprule
Comp & Group            & Subgroup                                                                                                                   & Complexity       & Country            \\ \midrule
1    & Minerals         & \begin{tabular}[c]{@{}l@{}}Coal, iron and \\ other primary products \\ (wheat, meet, wool)\end{tabular}                    & -                & Australia          \\ \hline
2    & Industry         & Textiles, engineering, others                                                                                              & Low and medium   & San Marino         \\ \hline
3    & Industry         & Vehicles and parts                                                                                                         & Medium           & Belgium            \\ \hline
4    & Industry         & Footwear, clothing and toys.                                                                                               & Low              & Macao              \\ \hline
5    & Industry         & \begin{tabular}[c]{@{}l@{}}Non-digital electronics, \\ record tapes,\\ telephone lines, \\ photographic paper\end{tabular} & High (up to 70') & Czechoslovakia     \\ \hline
6    & Industry         & \begin{tabular}[c]{@{}l@{}}Vehicles, boats,\\ machinery and parts\end{tabular}                                             & Medium and high  & Japan              \\ \hline
7    & Oil              & Petroleum gases                                                                                                            & -                & Turkmenistan       \\\hline
8    & Minerals         & Copper                                                                                                                     & -                & Chile              \\\hline
9    & Agricultural     & \begin{tabular}[c]{@{}l@{}}Coffee, bananas, other food\\ and primary products\end{tabular}                                 & -                & Reunion            \\\hline
10   & Industry         & Cars and electronics                                                                                                       & Medium and high  & Mexico             \\\hline
11   & Industry         & \begin{tabular}[c]{@{}l@{}}Cars, parts and\\ other machinery\end{tabular}                                                  & Medium           & Germany            \\\hline
12   & Oil              & Crude petroleum                                                                                                            & -                & South Sudan        \\\hline
13   & -                & Gold, watches, jewelry                                                                                                     & -                & Switzerland        \\\hline
14   & Industry         & \begin{tabular}[c]{@{}l@{}}Lubricating petroleum oils\\ and preparations\\ and other chemicals\end{tabular}                & -                & Curaçao            \\\hline
15   & Minerals         & Diamonds                                                                                                                   & -                & Botswana           \\\hline
16   & Industry + Agro. & \begin{tabular}[c]{@{}l@{}}Aircraft, auto parts,\\ soya and corn\end{tabular}                                              & Medium and high  & USA                \\\hline
17   & Industry + Agro. & \begin{tabular}[c]{@{}l@{}}Vehicles, parts,\\ wood and derivatives\end{tabular}                                            & Medium           & Finland            \\\hline
18   & Industry + Agro. & \begin{tabular}[c]{@{}l@{}}Primary Products\\ and textiles\end{tabular}                                                    & Low              & Christmas Island   \\\hline
19   & -                & \begin{tabular}[c]{@{}l@{}}Unclassified Special \\ transactions\end{tabular}                                               & -                & St. Maarten Island \\\hline
20   & Fuels            & Fuel oil, gasoil, etc.                                                                                                     & -                & Yemen              \\\hline
21   & Industry         & \begin{tabular}[c]{@{}l@{}}Medicaments,\\ medical appliances\\ and chemicals.\end{tabular}                                 & High             & Irlanda            \\\hline
22   & Oil + Agro       & \begin{tabular}[c]{@{}l@{}}Hydrocarbons,\\ palm oil, cocoa, etc.\end{tabular}                                              & -                & Ghana              \\\hline
23   & Industry         & \begin{tabular}[c]{@{}l@{}}Processors, \\ microcircuits, toys and shoes.\end{tabular}                                      & High and low     & China              \\\hline
24   & Industry + Agro. & Boats, meat, fish, dairy                                                                                                   & Medium           & Iceland            \\\hline
25   & Minerals + Agro. & Soya and derivatives, Iron                                                                                                 & -                & Paraguay           \\\hline
26   & Industry + Agro. & \begin{tabular}[c]{@{}l@{}}Aircrafts, vehicles,\\ perfumery, wine.\end{tabular}                                            & High             & France             \\\hline
27   & Industry         & \begin{tabular}[c]{@{}l@{}}Electronic microcircuits\\ and other machinery parts.\end{tabular}                              & High             & Philippines        \\\hline
28   & Industry + Agro. & Rice, cotton, textiles, gum, etc.                                                                                          & Low              & Pakistan           \\\hline
29   & Industry + Agro. & machinery, flowers, cheeses.                                                                                               & High             & Netherlands        \\\hline
30   & Industry         & Vehicles, parts and medicines                                                                                              & Medium and high  & United Kingdom     \\ \bottomrule
\end{tabular}
\caption{Description of the components. Groups, industrial complexity and country with greatest participation. K = 30.}
\label{table:desc_comp}
\end{table}

It is interesting to particularly highlight component 5, albeit (as mentioned below) it is not defined into a few products, given its high tech complexity\footnote{That is, recording tapes, telephone lines or photographic paper.} at the beginning of the series (during the '60s), but which later fell into disuse or decreased their share in international trade. In this sense, it is unsurprising that Czechoslovakia would be the most characteristic country of this component, given that, due to the country's dissolution in 1992, its time series is shorter than the rest.\footnote{This means that its denominator is lower than that of the other countries of the dataset.}

\subsection{Country Analysis per Component: some Stylised Facts}

Having the labelled components, this subsection analyses each country's exports basket composition  over the period under study (1962-2016). Since by definition our unit is country-year, it is possible to compare the evolution in components' distribution within each country. Below, we highlight some regularities that can be inferred from looking at the exports shares of the main country in each component.

The following analysis intends to present one of the various possible analysis that could be performed with the LDA application proposed in this paper. Rather than being exhaustive, the intention of this section is to present results in a way that allows to understand the possible derivations of the methodology. A previous time-series analysis for various country groups (i.e. covering some oil producing, North American, European, Latin American and Asian countries) focused its attention upon the evolution of those countries' exporting baskets. The overall conclusion is that the evolution of exports in Asia leaves a very different image to that of Europe. While in the latter the concentration of the EU countries’ export baskets in a single component shows a national differentiation, this is not the case in Asian countries, which show a homogenization process.

\begin{figure}[htbp]
	\centering
	\includegraphics[width=0.75\linewidth]{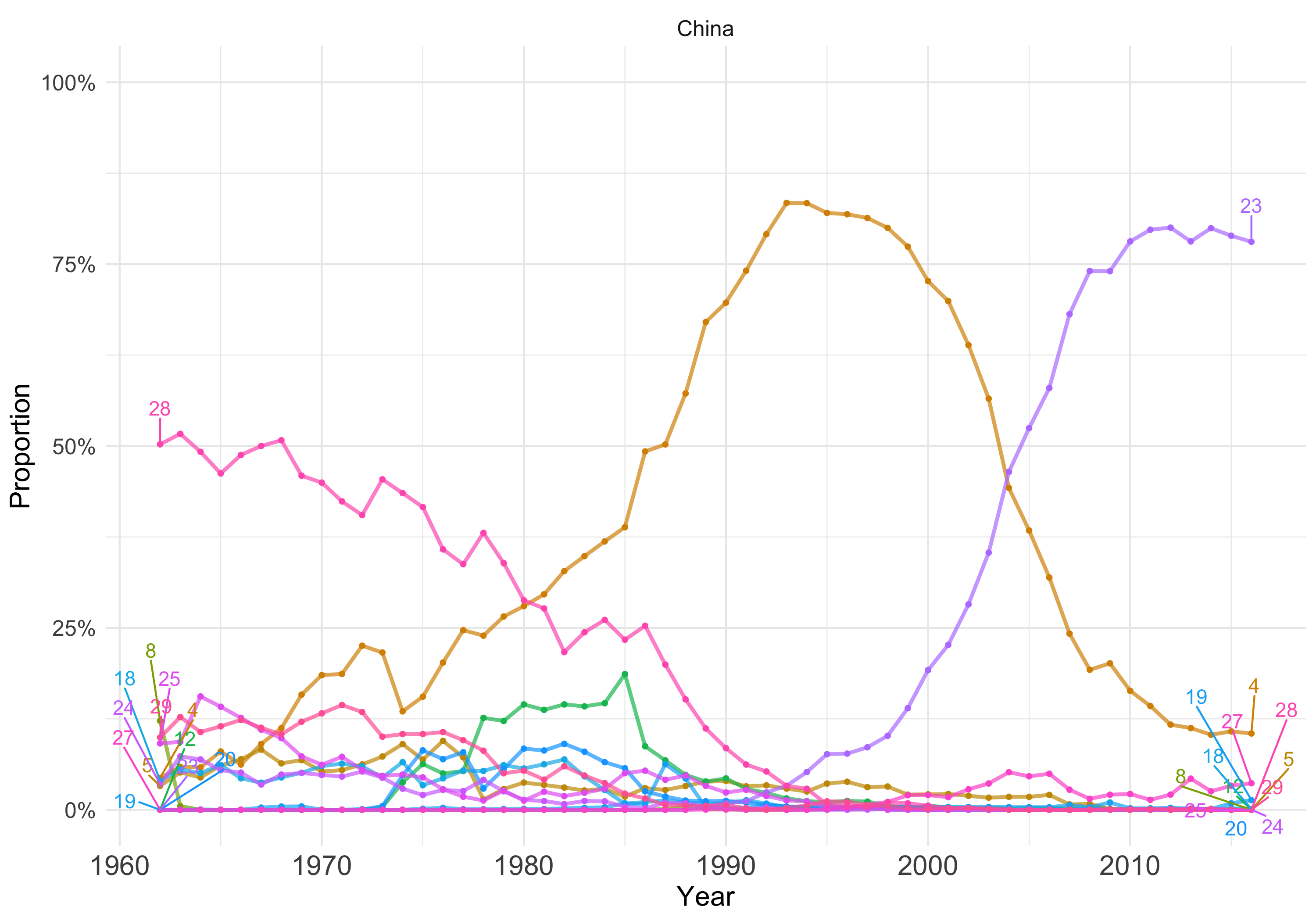}
	\caption{Distribution of components: China}
	\label{fig:graficoldak30chn}
\end{figure}

LDA results for Chinese exports structure show an interesting example to highlight (see Figure \ref{fig:graficoldak30chn}). At the beginning of the ‘60s the most relevant component (28) was composed of rice, cotton, tea and some textile products. This component shows a downward trend, while clothing, toys, etc. (component 4) increases and becomes the most important over the period 1980-2003. However, from 1993 textiles and toys start decreasing, with a simultaneous rise in component 23 (televisions, computers, microcircuits and transistors), which towards the last years of the period constitutes approximately 80\% of the country's exports. This change in the specific nature of Chinese exports reflects three stages of increasing complexity of the country's manufacturing industry, starting from a basically agricultural economy and, after a period of low-complexity industrialisation, becoming one of the world's leading exporters of highly complex products \parencite{chenery1986industrialization, costantino2013gatopardismo}.

Among other findings, \cite{Kozlowski2019thesis} highlights the concentration of EU countries' export baskets in a single component that varies between countries, hence showing national differentiation, while in most Asian countries tend to show much more homogeneous exports baskets. Further, it finds a clear concentration of the exports baskets of the Organization of the Petroleum Exporting Countries (OPEC) founding members, with some differences for Venezuela (mainly due to its particular history and some of its active public policies to diversify the countries exports \cite{bertola2010desenvolvimento}. Another result worth mentioning is the that our LDA model captures the exports specialisation in electronic products in the United States moving from analogue to digital technologies over the period of study, together with the Maquila phenomenon in Mexico.

The following is a summary of some general comments and regularities identified in the results for our LDA model, by looking at each component and with the aim of understanding what could such results be reflecting in terms of product composition (or exporting basket). To do so, we analyse the granularity and homogeneity of each component and confront its products with the export basket of the main country identified.

It is first comforting to find that 23 of the 30 components are relatively well defined (i.e. with the first ten 4-digits products explaining a cumulative probability over 30\%). And this includes one component (19) which groups, with a 96\% probability, unclassified commodities (i.e. “Special transactions, commodity not classified according to class”). In this sense, components 2, 16, 17, 18, 24 y 29 present a mixture of quite different products, making it difficult to understand an export pattern for a certain country, while component 5 is rather atomized (with only 18\% grouped in the main ten products).

In general terms, our LDA model seems to capture those countries with strong export basket concentration, either at the beginning or (and mostly) at the end of the period. In other words, those countries with an important growth of a certain product tend to be the main actor in the component that concentrates such product.

A brief characterisation of what our LDA model may be capturing over time can be divided into five groups, contrasting the key products with the main country's exports within each component. The 1962-2016 time series should be long enough to show important structural changes in each country's export basket.

On the one hand, ten (out of the mentioned 23) components show a main country with significant exports increases. First, Turkmenistan (the main country in component 7) “Petroleum gases, nes, in gaseous state” exports rose from 0.1\% (in 1995) to 73\% (in 2016), while Philippines' export share of “Electronic microcircuits” (component 27) grew from 0.01\% (in 1971) to become its first exporting good (with 27.2\% over total goods' exports in 2016). Other impressive increases are shown by Ireland (component 21, with “Medicaments (including veterinary medicaments)” going from 0.1\% to 15.2\%), and China, with exports of “Television, radio-broadcasting; transmitters, etc” (component 23) rising from null to 6\% (becoming its main export product, even including services, in 2016). Further, Australia (component 1) saw a rise in “other coal, not agglomerated” exports from 1\% (in 1962) to 14.1\% (2016), while looking at component 6, Japanese exports of “Passenger motor vehicles (excluding buses)” grew from 0.6\% (1962) to 13.9\% (2016). In component 10, Mexico's exports of "Passenger motor vehicles (excluding buses)" grew from 0\% to 8\% over the period. Also, french export shares of ``Aircraft of an unladen weight exceeding 15000 kg” (main product in component 26), went from 0.3\% to 8.6\%.\footnote{The remaining products in the component are not relevant in terms of the country´s export shares.} Moreover, in component 25, the second and third products are significant in terms of Paraguayan exports and show important rises: ``Oilcake and other residues (except dregs)” increased from 2\% to 12.3\% and sales of ``Soya beans” from 0.3\% in 1963 to 23\% (becoming the country's main exporting product, even including services, in 2016). Moreover, British exports of ``Passenger motor vehicles (excluding buses)” remained practically stable (5.2 to 5.3\%), although the following relevant products in component 30 (``Parts, nes of the aircraft of heading 792” and ``Medicaments (including veterinary medicaments)”) saw significant increases (from 0.3\% to 3.6\%, and from 0.6\% to 5.3\%, respectively).

A second group of (three) components shows significant falls over the period. In component 8, where Chilean exports of “Copper and copper alloys, refined or not, unwrought” fell from 30.3\% in 1962 to 22.6\% in 2016, while “Copper ore and concentrates; copper matte; cement copper” exports decreased from 33.1\% to 19.1\%. Also, Finish (component 17) “Wood of coniferous species, sawn, planed, tongued, grooved, etc” exports fell from 21.4\% (1962) to 2.7\% (2016), while Pakistan (component 28) saw a shrinking share of its “Raw cotton, excluding linters, not carded or combed” exports, from 9.8\% to 0.2\%.\footnote{This country's exports are stable in the other two main products of the component: “Rice, semi-milled or wholly milled” (6\% to 7.3\%) and very low in “Precious jewellery, goldsmiths' or silversmiths' wares” (0\% to 0.2\%).}

Another result worth highlighting shows one component with a mixture of the second or third main products (which still have a probability similar to that of the first one) with significant exports both falls and increases. In component 13, Switzerland's exports of “Watches, watch movements and case” fell from 12.7\% (in 1962) to 6.5\% (in 2016), but “Gold, non-monetary (excluding gold ores and concentrates)” rose from 0\% to 28\% over the same period.

A fourth group is formed by two components that show relatively constant trade over the period. In component 4, Macao experienced stable “Footwear” exports (from 4.1\% in 1962 to 3.9\% in 2016), and hence its emergence can probably be explained by its significant share in services exports (with tourism taking 88.8\%).\footnote{Other products within this component do not seem to be relevant in the country's exports basket: “Children's toys, indoor games, etc” fall from 0.6\% to 0.1\% and “Outerwear knitted or crocheted, not elastic nor rubberized; jerseys, pullovers, slip-overs, cardigans, etc” from 0.4\% to 0.1\% over the same period.} On the other hand, Germany (in component 11) exported an 8\% (in 1962) and 11.2\% (in 2016) in “Passenger motor vehicles (excluding buses)”, although its preponderance can be due to the fact that it is the main world exporter of this good (see below).

A singularity of this LDA trade data application is that in some (five) components it singles out countries with a short time series due to their shorter data history, as mentioned to explain Czechoslovakia in component 5 in the previous subsection. This is the case of the aforementioned (component 7) Turkmenistan (with data from 1992), while Réunion data ranges over the 1962-1995 period and it mainly exports “Sugars, beet and cane, raw, solid” (third main product from component 9, with a 4\% probability), with its exports basket shows an important concentration of this product (albeit falling from 83.6\% to 66.2\%). South Sudan (main country in component 12) exported 98.7\% in “Crude petroleum and oils obtained from bituminous materials” in 2016, but it only presents data from 2012, while Kuwait was the main exporting country of this product in 1962 (albeit falling from 17.7\% to 7.5\% in 2016) and Saudi Arabia in 2016 (rising from 10.5\% in 1962 to 18.2\%). Also, Curaçao (component 14) presents data only for 2011-2016 and Botswana (component 15) from 2000 (with 64.3\% probability in ``Diamonds (non-industrial), not mounted or set” exports and rising to 88.3\% in 2016), albeit the country only exported 1.4\% of that product globally in 2000 (although that share grew to 4.7\% in 2016).\footnote{The UK and India were, respectively, the main exporters of this product in 1962 and 2016. This result may be reflecting Botswana's relative comparative advantage in diamonds (i.e. a large share within its exports basket vis-à-vis the world average).}

Further, only one component (22) does not show a particular regularity that can explain the representative country (Ghana): its main product (“Petroleum gases and other gaseous hydrocarbons, nes, liquefied”, with a 38\% probability) is currently mainly exported by Qatar, rising from 0.2\% (in 1975) to 22.4\% (in 2016).\footnote{Conversely, Ghana's exports rose in “Palm oil” (with 10\% probability in the component) but from 0.01\% to 0.6\%, “Natural rubber latex; natural rubber and gums” (6\%), from 0.1\% to 0.2\%, and “Cocoa butter and paste” (1\%; from 2.8\% to 4.4\%). Over the period, Ghanaian exports fell in “Sawlogs and veneer logs, of non-coniferous species” (4\%) from 7.6\% to 0.9\% and “Cocoa beans, raw, roasted” (3\%) from 59.8\% to 16.9\%, “Wood, non-coniferous species, sawn, planed, tongued, grooved, etc” (3\%) from  6.1\% to 0.9\%, “Plywood consisting solely of sheets of wood” (3\%) from 0.4\% to 0.02\%, and remained stable in “Tin and tin alloys, unwrought” (2\%) and “Palm kernel oil” (1\%), both with null (or almost null) exports.}

Finally, another interesting fact derived from our LDA model is that there is one product (“Passenger motor vehicles (excluding buses)”) captured as the main one in six of the 30 components (3, 6, 10, 11,  17 and 30). This seems to reflect different exports baskets specialisation in the main country for each component (respectively, Belgium, Japan, Mexico, Germany, Finland and UK). As previously mentioned, Germany (component 11) has been the main exporter of this product over the whole period (albeit with a falling share from 37.6\% to 22.1\% over total exports), while the Japanese share (component 6) grew from 1.9\% to 13.5\%, those from UK and Belgium fell (from 19.6\% to 5.9\%, component 30; and from 4.7\% to 3.8\%, component 6; respectively), Mexico's rose (from 0\% to 4.7\%; component 10), and Finland's was the lowest (from 0.1\% a 1.8\%; component 17).

\section{Discussion}
\label{sec:discussion}

The present work proposes the use of a technique widely explored in Natural Language Processing to the field of international trade. By shifting the data domain from text to each country’s exports flows of each product, we managed to develop a typology of global trade based on a number of latent components. This allows us to do two things. On the one hand, we build an automatic classification of products based on data. On the other, we are able to study the trends in countries’ exports, based on those components. Our findings are mostly in line with the specialized literature for each country or region, showing that this particular methodology is able to grasp an insight of the position of countries’s exports in global trade, making use of a single type of metric.
Given that this methodology requires a minimum number of arbitrary decisions to be built, it turns out to be an interesting complement to the traditional forms of analysis.

The limit of the proposed methodology is its dependence of the data inputs. Decisions made with respect of the curation of the dataset can potentially affect all the results. If the dataset used starts in the beginning of the $20^{th}$ century, the resultant components would be very different with the ones presented in the article due to the larger set of technologies involved, and the optimal number of components would probably increase. On the other hand, if a country is restricted to a subset of the years considered, it will have a overall closer relation with components specialized in technologies of that time-frame, like in the case of Czechoslovakia. Even when each country-year weights the same in the optimization of the model, i.e. we are not considering the weight of the total exports of each country-year on the cost function, countries with larger exports tend to show smoother results, as is the case of China. This is due to the fact that the higher exports make it difficult for a specific product to drastically change its proportion in the total exports of the country from one year to another. Small countries are prone to sudden changes in the proportion of components, because a small change in the nominal value of the exports of any specific product imply a big proportion over the total basket of exports.
There is also an interesting phenomena that occurs on the model with countries that have a highly concentrated export basket. For the OPEC countries we can see a drastic change by the end of the 70'. If we take the case of Iraq, for example, it goes from an equal distribution on components 20 and 12 to a 100\% in the component 12 some years later. The distribution on the original SITC classification shows that this country exported 61.68\% in "Crude petroleum" and 36.5\% "Petroleum products, refined" in 1977, and the next year this changed to 85.03\% and 12.59\% respectively. This imply an increase of more than 23\% of the overall basket in a single product. Still, it is not a 50\% change as showed by the proposed model. The explanation for this is that both latent components, 12 and 20 include, with different proportions, crude and refined petroleum. The model infer that the refined petroleum exported from the 1978 on-wards comes from a different latent component than the one exported previously. We can say that if a countries export can be correctly describe only with two products, like in this case, using a model like LDA is not necessary for studying the exports basket.
Another interesting phenomena that this model cannot fully capture is the case when the bilateral interactions imply both imports and exports of highly complex product, and where one of the poles only produce a simple step in the production, like the mentioned Mexican maquilas. As we only use exports data, the model can only account for half of the process, producing potentially misleading conclusion if not used carefully. This problem, however, will arise in every metric that only accounts for the exports.

Benchmarking the results of the LDA model is a complicated task, as it is an unsupervised model. The best model should be the one that gives the most interpretable results, and that can be used for the more insightful analysis. To test our model, we tried three other approaches for the same task: finding the latent dimensions of international trade. First, we try two other methods traditionally used for Topic Modeling in Natural Language Processing, namely Latent Semantic Analysis (LSA) \parencite{landauer2013handbook} and Non-Negative Matrix Factorization (NMF) \parencite{lee1999learning}. Then, we tried to adapt the product space \cite{Hidalgo2007,Hidalgo2009} to achieve the same task as LDA, by using clustering techniques \parencite{kaufman1987clustering}. The three techniques showed results that are in line with the ones found by LDA, but in a lower level of detail, where the interpretation of results became a harder task.

It is interesting to look at the feasibility of the model given the change in the domain of the problem. The very different nature of the data traditionally used in text mining and Topic Modelling, with respect to international trade data, raises the question of whether the model can operate in the new domain. However, in terms of data structure, both problems have more similarities than what it seems. First, the traditional dimension of the problem is NxV (N observations, in the order of magnitude of thousands, V the vocabulary, also in the order of magnitude of thousands). In this case, the problem is approximately NxP, where the N observations are the year-country pairs, with 250 countries and 54 years, and P products, which in SITC at 4 digits are approximately 750. In other words, we are in an order of magnitude similar to that of a small dataset in a traditional Topic Modelling problem. Finally, an important change in both domains is the difference between the frequency of words in a text (tens or hundreds, depending on size of the documents) and the dollars exported of a product by each country-year (millions or billions). This difference in principle should not affect the model, since what the model considers in its optimization are the distributions between the different elements (word frequencies or exported values per product) and not the absolute values.

As future lines of work, as results are deeply connected with the input dataset, new data sources could provide different insights. For example, while our period seems long enough to reflect structural changes, economic historians could find an even longer time series more useful to describe some phenomena. Also, including services to the dataset could show different aspects of global trade that cannot be captured in an analysis only covering trade in goods. That said, data limitations would pose a trade-off, as this would imply either a lower product dissagregation or a shorter time series dataset. Other lines of work involve an exploration by country groups, in order for example to explore specialisation or complementarity among countries exports baskets, e.g. within a regional trade block.

As final remark, we do not think this new types of techniques will be able to replace traditional metrics and empirical work on international trade, but rather we intend to complement traditional analysis and bring a new tool that might help in the understanding of this field.

 \section*{Conflict of interest}
 The authors declare that they have no conflict of interest.
 
\section*{Acknowledgment}
The Doctoral Training Unit Data-driven computational modelling and applications (DRIVEN) is funded by the Luxembourg National Research Fund under the PRIDE programme (PRIDE17/12252781), \url{https://driven.uni.lu}.

This research was partly founded by the Préstamo BID - Proyecto de Investigación Científica y Tecnológica (PICT) 2016-1185.

Authors would like to acknowledge useful discussion with Daniel Heymann, Daniel Aromí and Jun Pang.

\printbibliography

%\end{document}
% end of file template.tex
\newpage
\begin{appendices}

\appendix
\section{Model with k=2}\label{appendix:k2}

\begin{figure}[h]
	\centering
	\subfigure[First component]{\label{fig:top_dist1}\includegraphics[width=\linewidth]{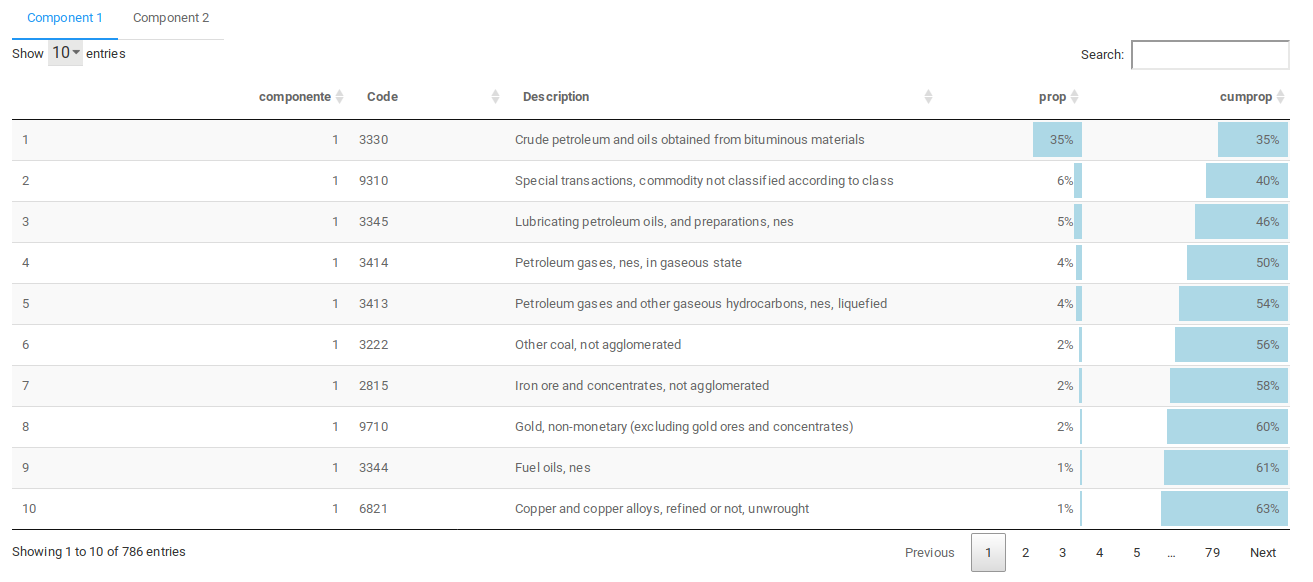}}
	\subfigure[Second component]{\label{fig:top_dist2}\includegraphics[width=\linewidth]{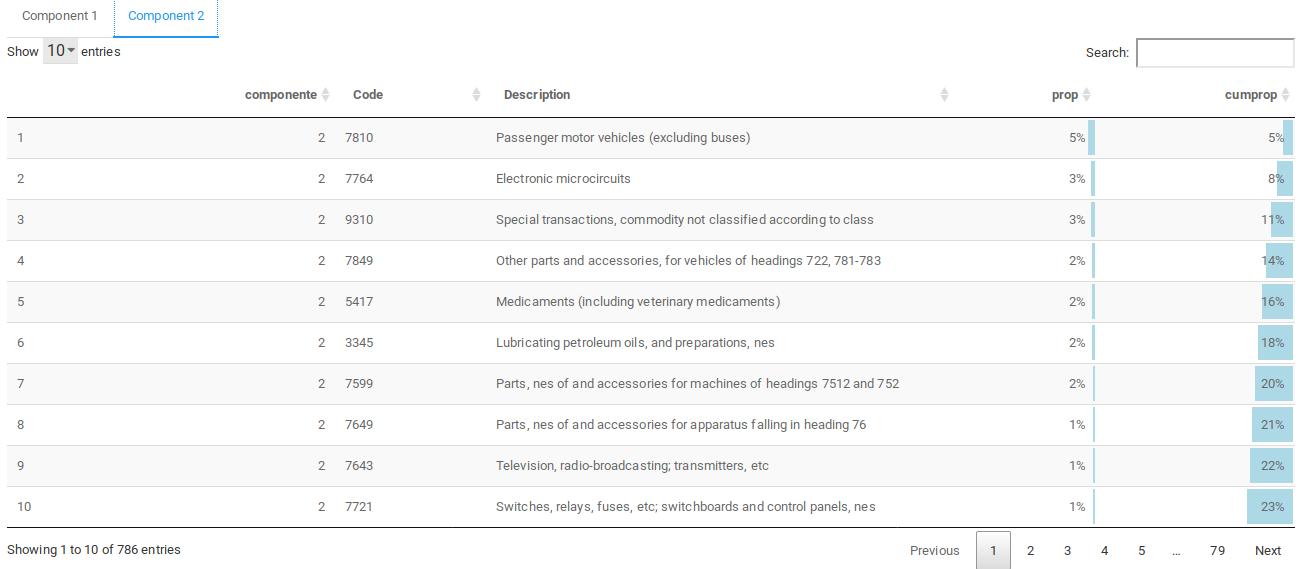}}
	\caption{screenshot of the interface for component characterization Highlighting of the proportion of the product in the component, and cumulative distribution. k=2}
	\label{fig:top_dist}
\end{figure}

Figure \ref{fig:top_dist} displays the mentioned interface in the case of $k=2$, showing each 4-digit SITC (Rev. 2) code and its product description, together with its individual and accumulated probabilities within the component. Further, Figure \ref{fig:top_dist1} shows that the distribution of the first component  assigns a large weight to crude oil, followed by other petroleum products (e.g. diesel oil, propane gas, etc.). Hence, a plausible label for such component would be "Petroleum and derivatives". However, it is also worth noting that component 1 also holds other products such as coal and metals (e.g. iron, gold and copper). Figure \ref{fig:top_dist2} shows the distribution of the second component (with $k=2$), which is more homogeneous than the first component, as the first product weighs only 5 \%, and the most outstanding products are passenger vehicles, electronic microcircuits, parts and accessories, etc. Hence, this component can be labelled to represent manufactured products in general.

\begin{figure}[h]
	\centering
	\subfigure[First component]{\label{fig:lall_1}\includegraphics[width=0.45\linewidth]{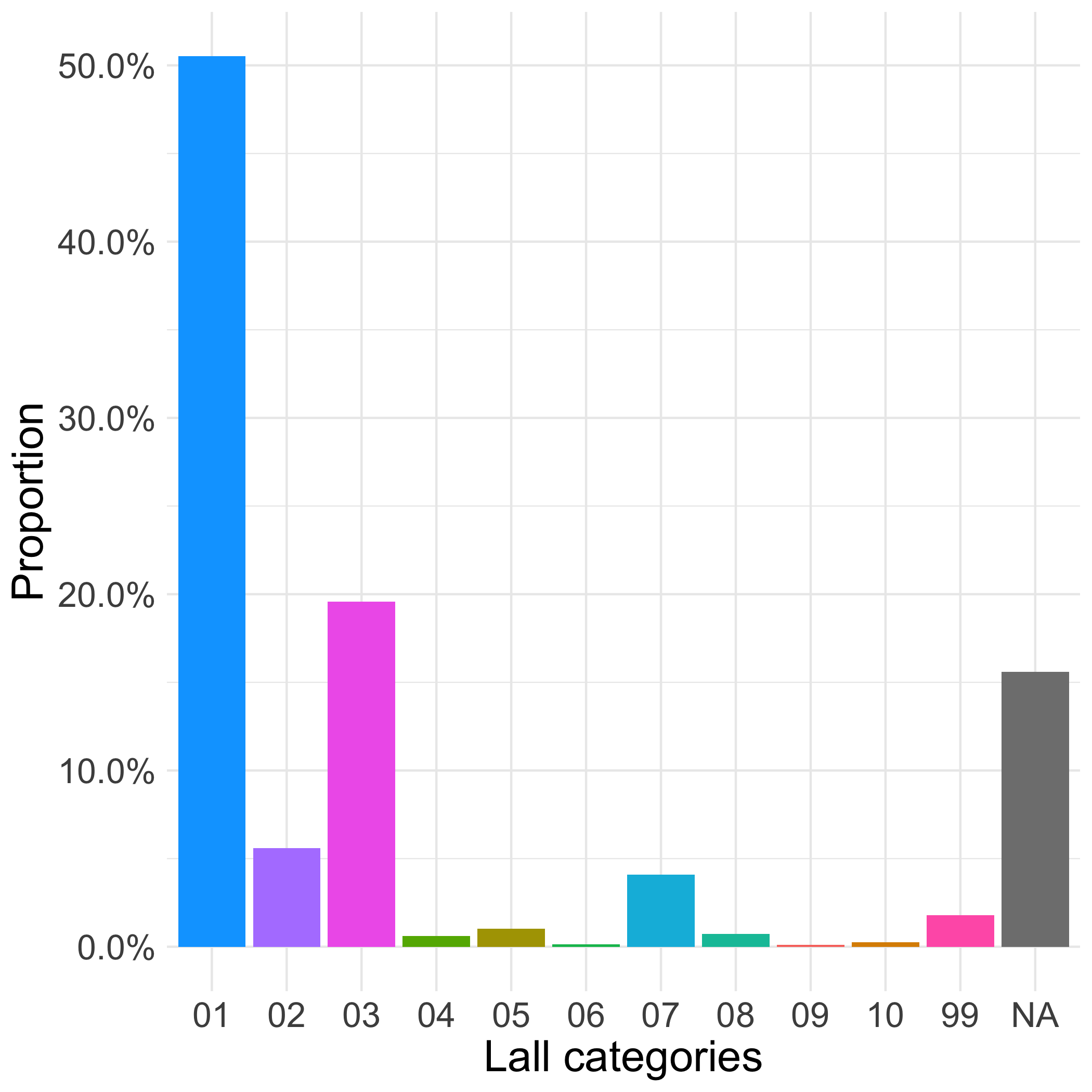}}
	\subfigure[Second component]{\label{fig:lall_2}\includegraphics[width=0.45\linewidth]{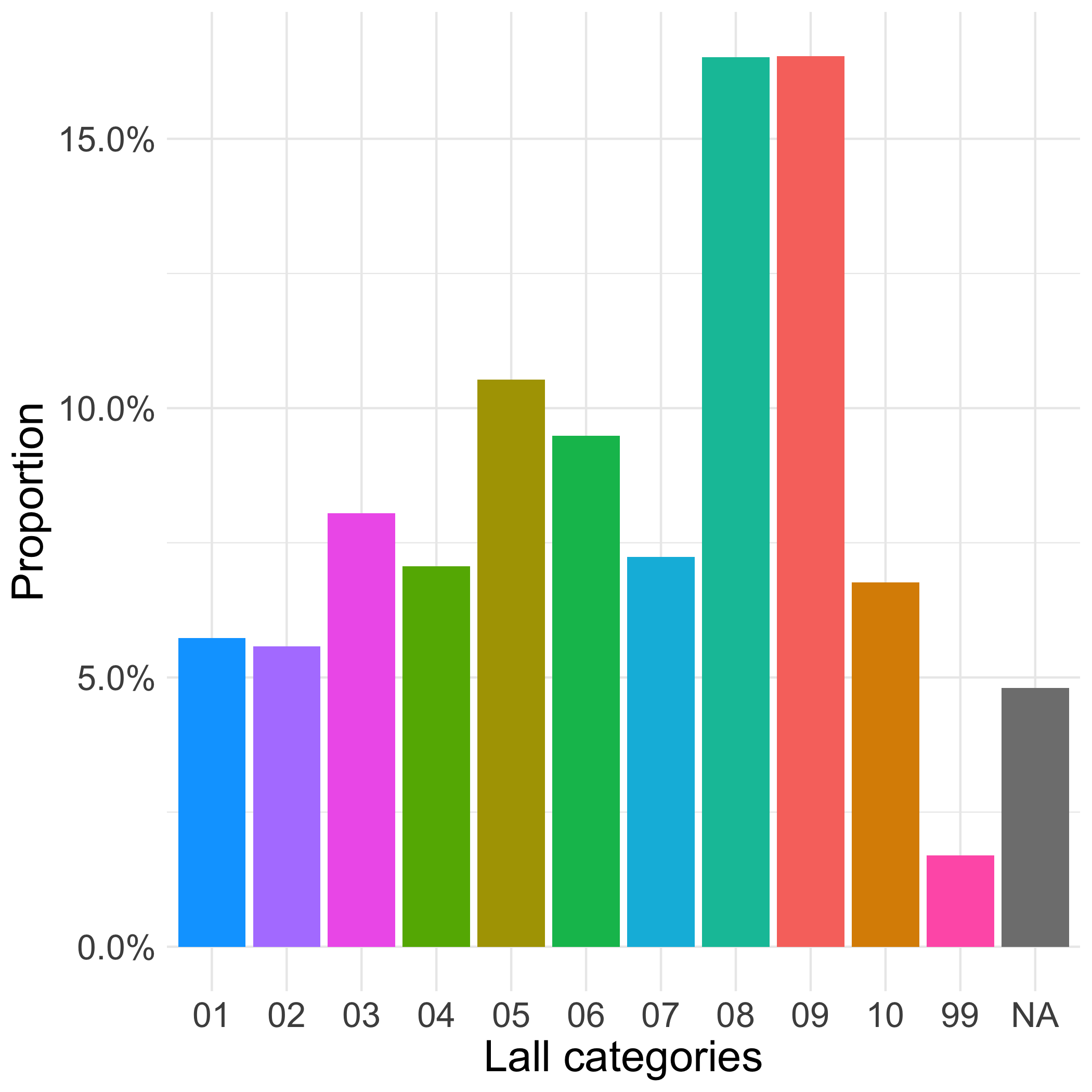}}
	\caption{Distribution of components by \cite{lall2000technological}. k=2. 01: primary products; 02: agro-resource-based manufactures; 03: non-agro-resource-based manufactures, 04: textile, garment and footwear (low-tech manufactures); 05:	other low-tech manufactures; 06: automotive (medium-tech manufactures); 07: process (medium-tech manufactures); 08:	engineering (medium-tech manufactures); 09: electronic and electrical (high-tech manufactures); 10: other high-tech manufactures; 99: unclassified products.}
	\label{fig:lall}
\end{figure}

Moreover, Figure \ref{fig:lall} shows the components' distribution (for $k=2$) according to the mentioned classification developed by \cite{lall2000technological}. According to that Figure, the first component is essentially composed of primary products and manufactures that use primary products as inputs. On the other hand, component 2 presents a more uniform distribution, where medium and high technology manufactures (e.g. engineering and electronics) stand out.

However, it is worth noting that for $k=2$, agricultural, livestock and forestry products cannot be singled out in one same component. That said, an interesting finding is that the division of the product space in only two groups allows the LDA model to find a first component mainly formed by petroleum (and its derivatives) products, while the other holds mostly manufactured products (SITC 5-8). In this sense, such model could allow understanding the classic corollary of comparative advantage models, where developed countries export manufactures (i.e. component 2) while developing countries specialise their trade in raw materials \parencite{balassa1979trade}. Some of the literature places a particular role to oil production (and exports) within an economy's structure \parencite{ross2012oil,carrera2017renta}. In this sense, with $k=2$  oil-producing countries' exports seem to lead the LDA model in finding its optimum by building one of the two components with such products. However, this dichotomy should be taken with care in the case of petroleum. As \parencite{ross2012oil} states, the resource curse of oil producing countries may be biased upward in poorer countries when using their dependence on hydrocarbon exports and derive "spurious associations between oil export dependence and a variety of economic and political maladies that are highly correlated with low incomes". This is hence an arguable statement, as oil exports reflect an indirect measure of a country’s non-oil economic size, although also the so-called "Dutch Disease" in oil-exporting countries has often crowded out their agricultural and manufacturing exports due to the cited comparative advantage \parencite{ross2012oil}.
\end{appendices}

\end{document}